# Electrokinetic Janus micromotors moving on topographically flat chemical patterns

Tao Huang[1,2], Vyacheslav R. Misko[3,4], Anja Caspari[5], Alla Synytska[5,6,7], Bergoi Ibarlucea[2], Franco Nori[4,8], Jürgen Fassbender[9], Gianaurelio Cuniberti[2], Denys Makarov[9], Larysa Baraban[1]

[1]*Helmholtz-Zentrum Dresden-Rossendorf e.V., Institute of Radiopharmaceutical Cancer Research, Bautzner Landstrasse 400, 01328 Dresden, Germany*
[2]*Max Bergmann Center of Biomaterials and Institute for Materials Science, Technische Universität Dresden, 01062 Dresden, Germany*
[3]*µFlow group, Department of Bioengineering Sciences, Department of Chemical Engineering, Vrije Universiteit Brussel, Pleinlaan 2, 1050 Brussels, Belgium*
[4]*Theoretical Quantum Physics Laboratory, RIKEN Cluster for Pioneering Research, Wako-shi, Saitama 351-0198, Japan*
[5]*Leibniz-Institut für Polymerforschung Dresden e.V., Hohe Straße 6, 01069, Dresden, Germany.*
[6]*Institute of Physical Chemistry and Polymer Physics, Technische Universität Dresden, 01062 Dresden, Germany*
[7]*Bavarian Polymer Institute (BPI), University of Bayreuth, Universitätsstraße 30, 95447 Bayreuth, Germany*
[8]*Physics Department, University of Michigan, Ann Arbor, Michigan 48109-1040, USA*
[9]*Helmholtz-Zentrum Dresden-Rossendorf e.V., Institute of Ion Beam Physics and Materials Research, Bautzner Landstrasse 400, 01328 Dresden, Germany*

**Abstract**. Ionic and molecular selectivity is considered unique for the nanoscale and not realizable in microfluidics. This is due to the scale-matching problem – a difficulty to match the dimensions of ions and electrostatic potential screening lengths with the micron-sized confinements. Here, we demonstrate a microscale realization of the ionic transport processes closely resembling those specific to ionic channels or in nanofluidic junctions, including selectivity, guidance through complex geometries and flow focusing. As a model system, we explore electrokinetic spherical Janus micromotors moving over charged surfaces with a complex spatial charge distribution and without any topographical wall. We discuss peculiarities of the long-range electrostatic interaction on the behavior of the system including interface crossing and reflection of positively charged particles from negatively charged interfaces. These results are crucial for understanding the electrokinetic transport of biochemical species under confinement, have the potential to increase the precision of lab-on-chip-based assays, as well as broadening use cases and control strategies of nano-/micromachinery.

**Keywords:** Janus micromotors, electrokinetic micromotors, ionic transport selectivity, microfluidics, nanofluidics, chemical pattern, topographically flat pattern

## Introduction

The transport of ions across macromolecular pores in cell membranes, the so-called ion channels, is essential to maintain life processes. Biological ion channels allow only specific species to cross the membrane with very high selectivity (Fig. 1A) (*1*). This functionality is broadly explored in nanofluidics for the realization of 'artificial ion channels' (*2*). Bioinspired nanofluidic systems consisting of nanopores, nanotubes, nanofabricated junctions (*3*), diodes and transistors enabled major discoveries in understanding the properties of fluids under nanoscale confinement and opened a new way to handle molecular or ionic species in fluids (*2*) for water purification, development of new generation of batteries, and processing of complex biological solutions, to name just a few (Fig. 1B) (*4, 5*).

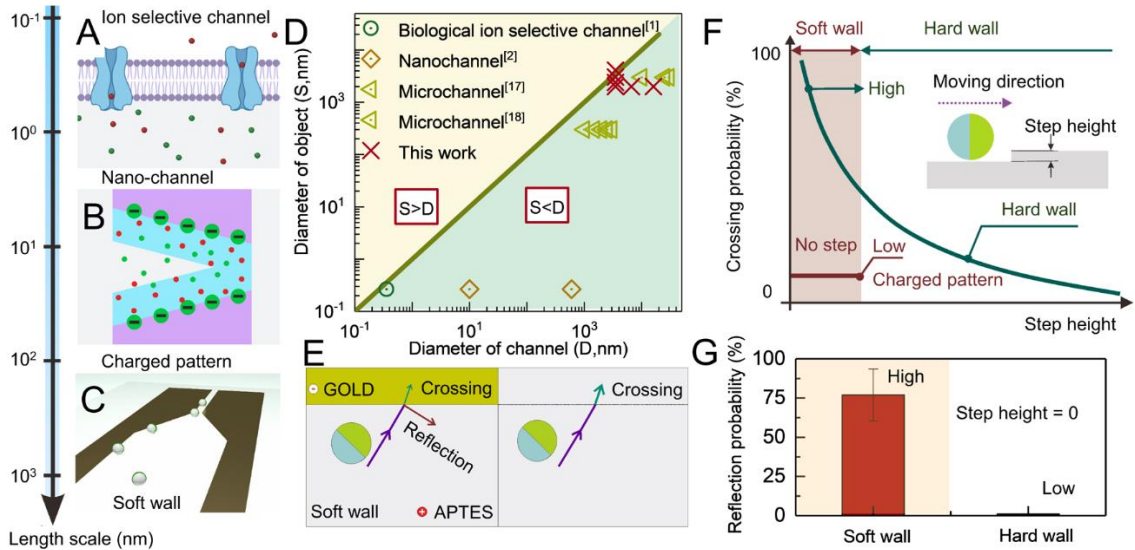

**Figure 1.** (**A**) Ions selectively across the biological cell membrane. (**B**) Ion sieving in a rectifying single nano-channel. (**C**) Schematic of macro-ions transport at micrometer scale relying on long-range interaction with soft charged walls. (**D**) Characteristic dimensions of channels (D) and objects (S) at different scales. Even when S>D, charged Janus particles moving in a potential with soft walls can pass through a narrow constriction in the pattern. (**E-G**) In the case of mechanical hard wall, the decrease of the height of the wall increases the probability of a micromotor to cross the wall (green line in panel (**F**)). On geometrically flat yet charged soft walls, charged Janus micromotors demonstrate a significant decrease of the crossing probability (red line in panel (**F**)). Instead of crossing the flat boundary, the majority of Janus particles are reflected from the flat charged interface (**G**).

Although ionic selectivity is quite common in multiple systems at the micrometer scale (e.g., nano- and micromachinery (*6*), and lab on chip bioassays (*7-9*) ), the effect is considered to be specific to nanoscale and can barely be reproduced in microfluidics. This is due to the difficulties in matching the dimensions of ions and electrostatic potential screening lengths with the micron-sized confinement (Fig. 1D). The geometrical mismatch is huge when we compare the motion of ions in micrometer large channels with their biological counterparts, where the charge selectivity is due to hydration processes and steric interactions with characteristic lengths of about 1-5 nm (*1*). There are orders of magnitude difference even when compared to nanofluidics, where the mechanism for charge selectivity is due to van der Waals or electrostatic interactions between a charged wall and ions (characteristic screening length of about 50-100 nm).

To overcome this scale-matching problem and to emulate the ion transport phenomena in microfluidics, we propose a macroscopic model of a motile ion, the so-called macro-ion. In this scenario, functional motile macro-ions should be placed and guided in microchannels relying on long-range electrostatic interaction. This task calls for the investigation of the interaction between the channel potential and macro-ions and propagation of the particles through microfabricated pores. Charged colloidal particles (*10, 11*) and spherical Janus micromotors (*12*) driven by the ionic self-electrophoresis (*13, 14*) can serve as a model system of macro-ions to carry out fundamental studies of the ionic transport at the microscale.

In this respect, Uspal et al. theoretically proposed to use chemically patterned surfaces for catalytic Janus particle and predicted a multitude of interaction scenarios, including reflection of particles from the boundary, hovering, and sliding along the boundary (*15-17*). Here we demonstrate ionic transport processes at the microscale by exploring electro-kinetic spherical Janus micromotors moving over charged surfaces with a complex spatial charge distribution. We realize *topographically flat* complex patterns on a chemically functionalized substrate containing regions of positive and negative Zeta potentials. We show that flat charged patterns enable long-range soft potentials – that smoothly change with the interaction distance– allowing to realize experimentally 'soft walls' that affect the motion of charged Janus colloids. Relying on the presence of these electrostatic potentials and exploring the ionic self-diffusiophoresis propulsion mechanism of Janus particles, we demonstrate the ability of the motile macro-ions to navigate in diverse chemical pattern geometries. Because the nature of the interactions between a soft wall and a macro-ion is distinct from the one characteristic for mechanical hard walls (Fig. 1, C and D) (*18-21*), we observe unconventional phenomena, when the particle approaches the wall (Fig. 1, E to G). Main differences are observed in processes related to the *reflection from-* or *crossing the wall* by a motile particle (*22, 23*). For instance, while a decrease in the height of a traditional 'hard wall' increases the probability of the micromotor to cross the wall (Fig. 1E, right side), the use of a geometrically flat soft wall demonstrates a large decrease of the crossing probability (Fig. 1E, left side). Instead of crossing the flat boundary, around 80% of *positively* charged Janus particles in our experiment are reflected from the flat *negatively* charged interface (Fig. 1, F and G). Noticeably, the average interaction time between Janus micromotors and the charged interface is found to be about 0.2 s, which is two orders of magnitude shorter than during the interaction with a hard wall.

We use soft walls to construct a microscale pore constriction, resembling selective ion channels by geometry, and demonstrate the critical differences of the microparticles propagations compared to the hard-wall. This includes size-dependent selectivity, flow focusing, and the possibility to move through the constriction for a Janus particle with a size larger than the width of the constriction. These results have strong implications for the fields of smart nano-/microscale machines and lab-on-chip-based assays.

**Results and Discussion**

**Positively charged light-activated Janus micromotors**. We fabricate photocatalytic blue light-activated Janus particles consisting of a 60-nm-thick AgCl film prepared on polystyrene (PS) microspheres with a diameter of about 2 µm. In contrast to prior reports on plasmonic Ag/AgCl Janus micromotors (*24, 25*), we utilize Ag/AgCl/PS Janus particles which are coated with β-FeOOH nanocrystals (*12*). Fig. 2A shows a scanning electron microscopy (SEM) image of a capped microsphere revealing a substantial surface roughness. The element mapping, performed using energy-dispersive X-ray spectroscopy (EDX), indicates the presence of AgCl at only one side of the PS bead while the signal from Fe is detected on the whole surface of the particle (Fig. 2B). These Ag/AgCl/β-FeOOH/PS Janus microspheres exhibit positive Zeta potentials ($\zeta = +27$ mV), that correspond to a positive electric charge distributed over the surface (*23, 26*).

Once suspended in deionized water and dropped onto the surface of glass, Janus particles quickly sediment and experience thermal fluctuations. When illuminating them with blue light, Janus particles are brought into motion (*12*), driven by several parallel photocatalytic processes. The light initiates the decomposition of AgCl and results in an asymmetric electric field around the Janus particle (*12, 13, 24, 25, 27*). Due to the positive charge of the particle, it sticks to a negatively charged substrate and moves along a positively charged substrate in the direction *towards the catalytic cap* (Fig. 2C) with a speed of about 30 μm/s. This behavior is in contrast with previous reports on negatively charged photocatalytic swimmers, where the preferred orientation of the motion is *away* from the cap (*12, 24*).

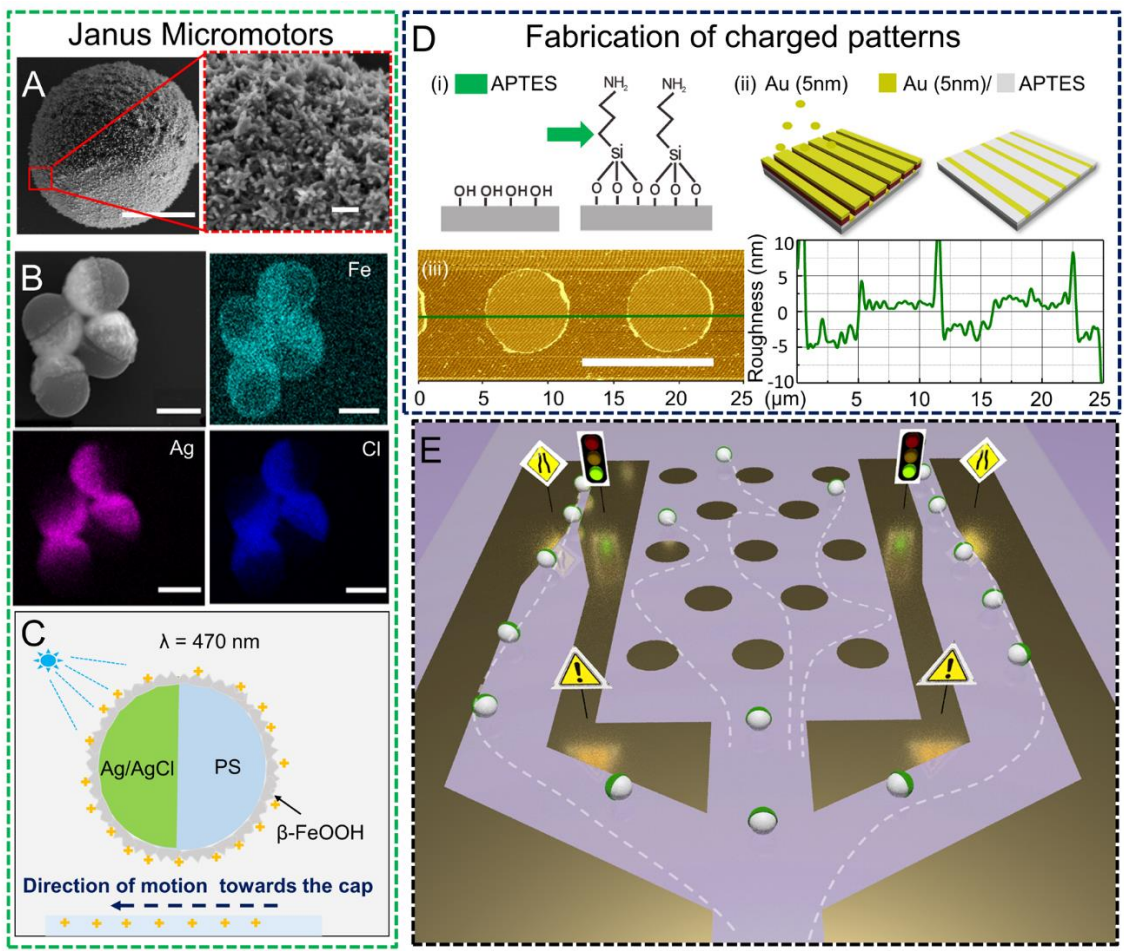

**Figure 2**. (**A**) SEM image of Ag/AgCl/ β-FeOOH/PS Janus micromotors, Scale bar: 1 μm. Inset: magnified SEM images of the PS-coated side of the particles. Scale bar: 200 nm. (**B**) EDX image of Janus micromotor for the Fe, Ag, Cl elements. Scale bar: 2 μm. (**C**) Schematic image illustrating the propulsion direction of the Ag/AgCl/β-FeOOH/PS Janus micromotor near a positively charged surface. Ag/AgCl/β-FeOOH/PS Janus micromotors move towards the cap. (**D**) Schematic image of the fabrication process and the charged pattern on a substrate. (iii) The Atomic Force Microscope (AFM) image of the APTES/Gold charged pattern (Left) and the corresponding surface roughness (Right). Scale bar: 5 μm. (**E**) Concept figure showing a Janus micromotor moving in a complex pattern with soft walls. Dashed lines correspond to the trajectories of Janus micromotors.

**Surfaces with complex spatial charge distribution.** We fabricate different topographically flat chemical patterns including stripes, circles, rings, and flow-focusing structures by deterministically tailoring the distribution of positive and negative charges at the surface (Fig. 1D). For this purpose, the self-assembled monolayer (SAM) of (3-Aminopropyl) triethoxysilane (APTES) was immobilized at the glass surface via the reaction of silanization. The Zeta potential of the APTES

SAM is $\zeta_{APTES}$ = +55 mV (pH = 5.65, Fig. S3). To achieve a negative charge distribution, and for better visualization of the pattern shapes and boundaries, we deposited a 5-nm-thick gold layer. The respective Zeta potentials for the gold layer is $\zeta_{Au}$ = −46 mV (pH = 5.65). The thickness of the gold layer is chosen to match that of the positively charged areas functionalized with APTES molecules (about 4 nm, Fig. 2D). Considering the diameter of Janus particles (2 µm) and their respective Brownian fluctuation scales (~ 1 µm), the separation distance between the bottom of the Janus particle and substrate is substantially larger than the 5 nm thickness of the gold layer. Furthermore, we estimate the Debye screening length to be about 200 nm and the particle-wall gap of several hundred nanometers (*12*). In this respect, we can safely neglect a slight variation between the thicknesses of the negatively and positively charged regions (about 1 nm difference) and consider the patterned surface as topographically flat. This renders the resulting chemically patterned substrates to be free of mechanically hard interfaces.

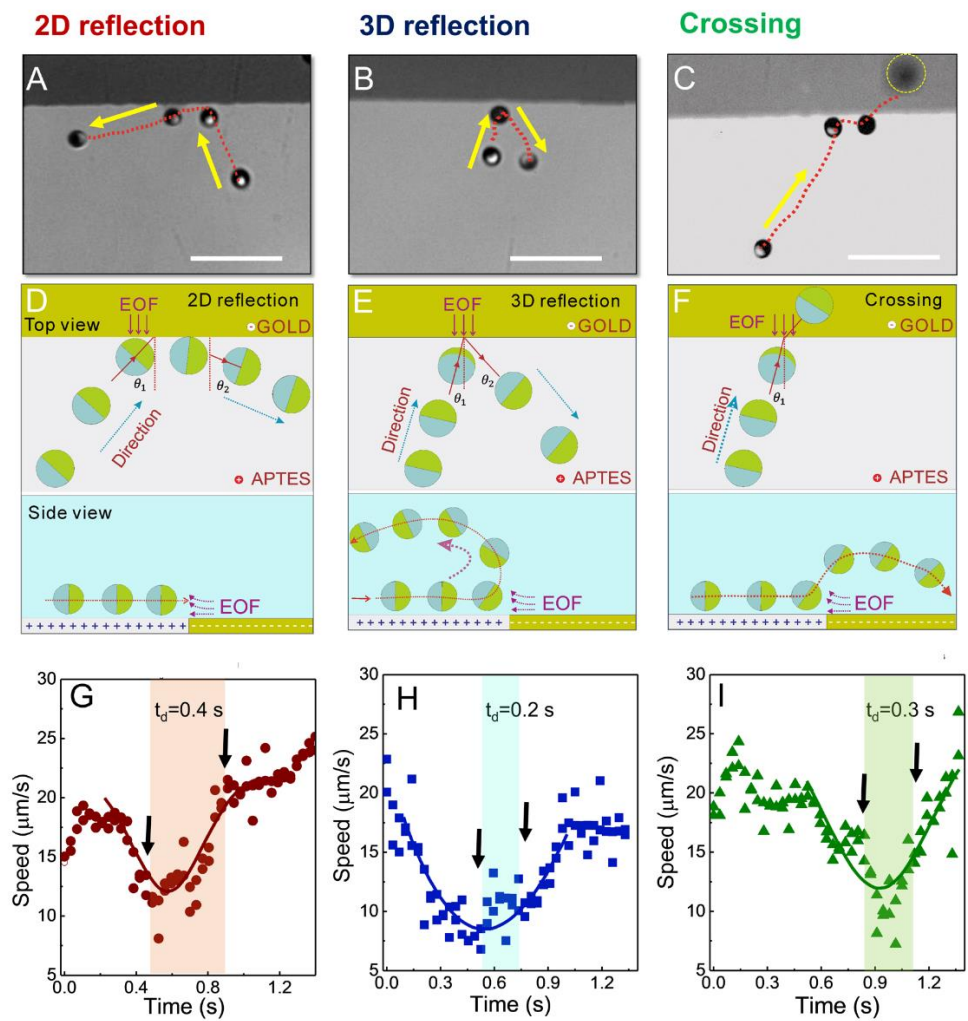

**Figure 3**. Positively charged Janus micromotors at the positive/negative interface. (**A-C**) When a Janus micromotor approaches the interface it can experience (**A**) 2D reflection, (**B**) 3D reflection, or (**C**) cross the interface. (**D-F**) Schematic illustration of the interaction of Janus particles with the interface: (**D**) 2D reflection, (**E**) 3D reflection, and (**F**) crossing the interface. The change of the speed of Janus micromotors during the process of (**G**) 2D reflection, (**H**) 3D reflection and (**I**) crossing the interface. The solid lines in G-I are guide to the eye.

**Janus macro-ions moving over charged surfaces.** When an aqueous suspension of positively charged Janus particles is placed onto the patterned substrate, the particles reveal Brownian motion when located above the APTES region but irreversibly stick to the negatively charged gold surface due to electrostatic attraction. Once the blue light illumination is switched on, Janus particles move along the APTES surface, propelling towards the cap. At the same time, the illumination does not bring in motion those Janus particles, which were stuck to the gold surface: they remain immobile. When motile positively charged particles approach the APTES/gold interface, most of the studied Janus particles (about 80%) experience *reflection from the interface of opposite charge*: positively charged particles are reflected away from the region with a negatively charged gold towards the positively charged APTES (Fig. 3, A and B). The remaining portion of the Janus particles (about 20%) *crosses the interface* and moves away from the substrate (Fig. 3C). Since Janus micromotors move towards their caps, the orientation of the particle cap is changed upon approaching the interface.

We distinguish two types of reflection of Janus particles from the APTES/gold interface: (i) two-dimensional (2D) rotation of the cap when the reorientation of the cap occurs within the substrate plane (Fig. 3, A and D; indicated as 2D reflection), and (ii) three-dimensional (3D) rotation of the cap when the particle reorients via an out-of-plane rotation of the cap (Fig. 3, B and E; indicated as 3D reflection). Fig. 3, D to F summarize the behavior of Janus beads at the interface, where the top view and side view of particles reflecting from the interface and crossing the interface are shown. Fig. 3D presents a sketch of the 2D reflection. The particle remains in the substrate plane and its axis reorients in the same plane via sliding along the interface and then detaching from it. In the case of 3D reflection, the Janus particle does not overcome the potential barrier at the APTES/gold interface. It is reflected back via lift and rotation out of the substrate plane (Fig. 3E). Finally, if the velocity of the particle is sufficiently high, it overcomes the potential barrier via lifting out of the substrate plane, and in this way crosses the interface (Fig. 3F).

We observe that the entire reorientation process is rather fast and, in most of cases, the persistence time, $t_b$, extends over several hundreds of milliseconds only (indicated bands in Fig. 3, G to I). Still, even this short interaction time of the particle with the interface is sufficient to assure that the reflection is not specular, with the incident angle ($\theta_1$) being smaller than the reflection angle ($\theta_2$) (Fig. 3, D and E). The corresponding statistics is summarized in Supporting Table S3. Remarkably, not only the orientation of the cap is affected but also the speed of Janus particles gradually decreases well before approaching the APTES/gold interface. The speed reaches its minimum at the interface and increases when moving back after the reflection (Fig. 3, G to I). Accompanied by the non-specular reflection from the interface, this slowing down suggests the presence of a long-range interaction potential between Janus micromotors driven by the ionic self-diffusiophoresis (*12, 13, 24, 25*) and electrostatic potential formed at the interface of the pattern. The experimental observation of the reflection of positively charged particles from the negatively charged gold surface indicates that the effect cannot be explained based on electrostatic interactions only. Additional factors related to the driving mechanism of the particles should be considered.

**Mechanism of reflection**. Light-activated Ag/AgCl/β-FeOOH/PS Janus micromotors release $Cl^-$ and $H^+$ ions and achieve self-propulsion due to the ionic self-diffusiophoresis mechanism (*12, 13, 28*). The difference in the diffusivity between $Cl^-$ and $H^+$ ions, results in a self-generated inward pointing electric field around the Janus micromotor. This field induces propulsion of Janus micromotors and electroosmotic flows (EOFs) along the surface of the charged substrate (*29, 30*). The EOFs induced by the self-generated electric field around the particle are illustrated in Fig. 4A. We simulated the electric potential distribution around the Janus particle and found a lowering of the electric potential near the AgCl cap. When a positively charged Janus particle moves towards the negatively charged gold interface with its cap forward, the self-generated electric field induces an EOF near the stationary charged interface (APTES and gold regions), as shown in Fig. 4, B, C.

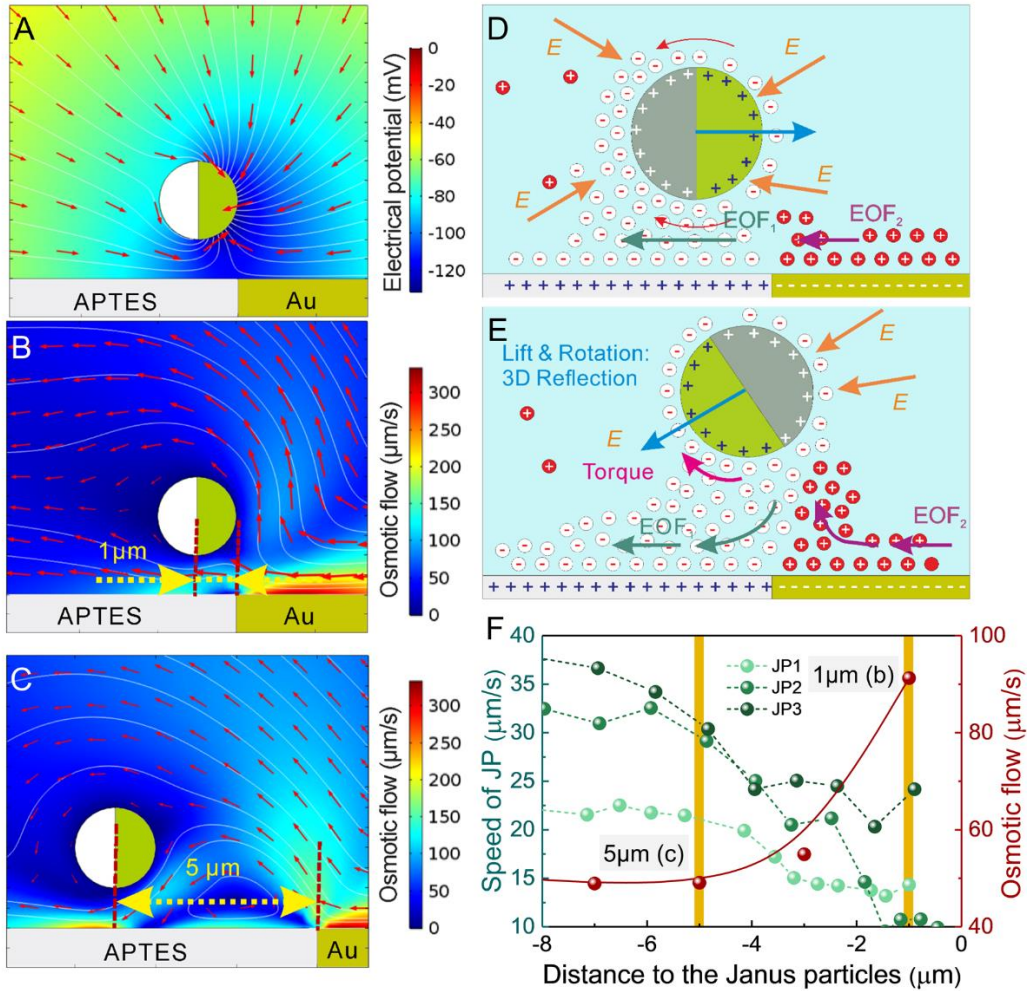

**Figure 4**. (A-C) COMSOL simulations of the electric potential and electroosmotic flow distributions around a Janus particle close to the interface between two oppositely charged surfaces. (A) The distribution of the electric potential (mV, color-coded) and electric field (red arrows) around a Janus particle. The distribution of the electroosmotic flow near the charged interface for the case when the separation distance between the center of the Janus particle and the APTES/Gold interface is (B) 1 μm and (C) 5 μm. The color code indicates the magnitude of the EOFs, and the red arrows indicate the flow direction of the EOFs. (D, E) Schematic illustrations of the mechanism of the repulsive interaction between a charged Janus micromotor and a charged interface. The electric field (E) generated by a Janus particle induced two EOFs near the positively charged APTES surface ($EOF_1$) and negatively charged gold surface ($EOF_2$). Both EOFs are very strong close to the charged pattern surface and flow in the same direction. EOFs create a torque that leads to the rotation of the Janus particle in the clockwise or counterclockwise direction. (F) Experimentally measured speed of three Janus particles (green line) and simulated speed of the electroosmotic flow under the Janus particle (dark red line) as a function of the separation distance between the Janus particle and APTES/Gold interface.

The direction of the EOF is determined by the sign of the surface charge and the direction of the electric field. As illustrated in Fig. 4, D and E, the surface charge near the positively charged APTES surface is formed by $Cl^-$ ions. These ions generate the EOF in the outward direction from the interface (green arrow in Fig. 4D). In contrast, at the gold side of the pattern, $H^+$ ions generate the EOFs towards the interface (shown by the dark red arrow in Fig. 4D). These two EOFs induced by oppositely charged ion species are responsible for the lift of the Janus particle during the 3D reflection (Fig. 4E). Furthermore, the EOFs decrease significantly when particles are far away from the charged substrate. Thus, the presence of EOFs creates a torque that leads to the rotation of the Janus particle in the clockwise or counterclockwise direction (Fig. 4E). As a result, the Janus particle lifts and the orientation of the cap flips backward, hence leading to a complex 3D reflection. Moreover, the magnitude of the EOF under the Janus particle (the green arrow in Fig. 4D) increases significantly, when the separation distance between the Janus particle and interface becomes smaller than 5 μm (Fig. 4F). Therefore, the specific profile of the EOF prevents the particle from moving forward and leads to the decrease of the velocity of Janus particles when approaching the interface. These simulation results are consistent with our experimental observations. In the case of the 2D reflection, when the separation distance between the Janus particle and interface is less than 5 μm, the speed of the Janus particle is found to decrease rapidly (Fig. 4F). The speed of a charged Janus particle is determined by a superposition of electrophoresis of the charged particle itself and the electro-osmotic flow caused by the charged wall. The speed of colloid micromotors undergoing ionic diffusiophoresis is proportional to the difference in Zeta potential between the colloid and the wall: $U \propto (\zeta_c - \zeta_w)$, where $\zeta_c$ and $\zeta_w$ are Zeta-potentials of colloidal micromotor and nearby charged wall (*12, 31-33*).

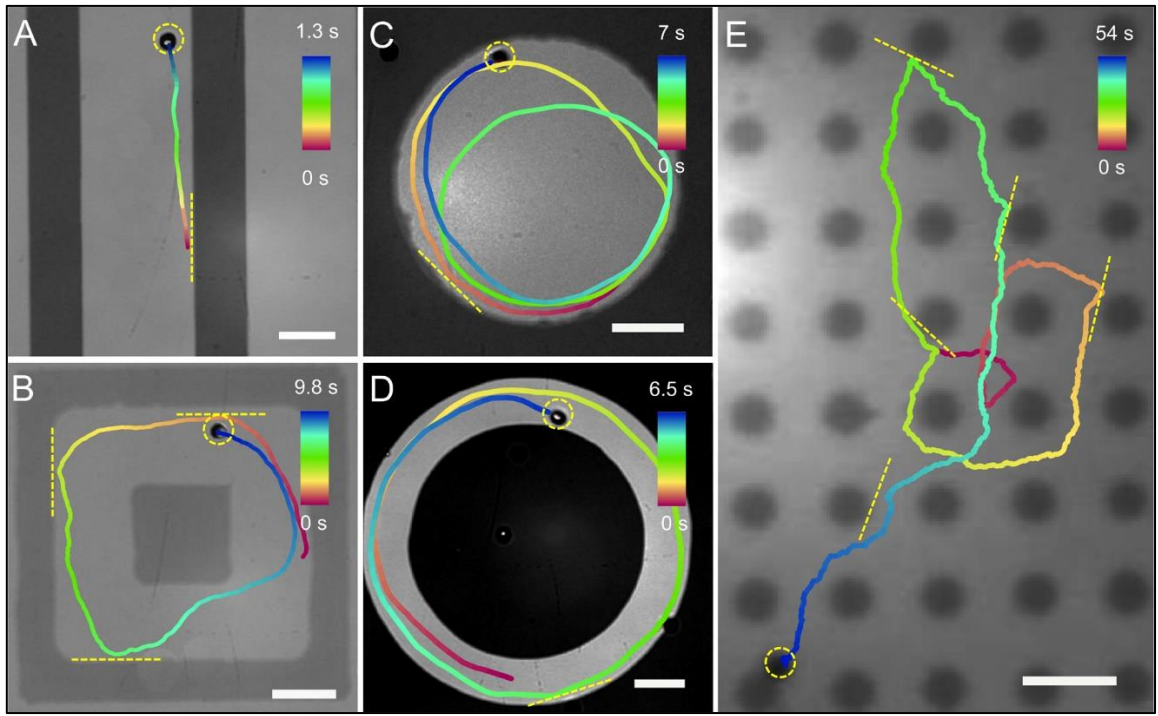

**Figure 5**. Navigation of Janus micromotors through various complex patterns. Optical micrograph of Ag/AgCl/β-FeOOH/PS Janus particles (labeled with a yellow dashed circle) on an APTES/Gold charged pattern. While the darker areas correspond to negatively charged gold, the brighter areas are positively charged APTES. The line with color mapping represents the trajectory of a Janus micromotor moving on (**A**) straight stripe pattern, (**B**) square ring, (**C**) circular pattern, (**D**) circle ring, and (**E**) circular dots array. The color-coded scale represents the observation time from the initial state (red) to the final state (blue). Scale bars: 10 μm.

**Navigation through patterns.** Understanding the interaction of charged Janus particles with the interface of positively/negatively charged surfaces can be used to navigate the macro-ions in complex microscale patterns, *e.g.*, circles, rings, and even achieve filtering of the transport through specially designed constrictions. The guidance is based on *multiple reflections* of Janus micromotor's from the interface, similarly to the propagation of light in optical waveguides. This behavior is distinct from the previously demonstrated cases of the particle's motion in a microfabricated patterned environment (*18, 19*), where micromotors prefer to follow the interface (persistence time $t_b \gg 1$ s).

Fig. 5 summarizes the experimental results demonstrating the guidance of Ag/AgCl/ β-FeOOH/PS Janus particles through the diverse pattern geometries, *e.g.*, linear channels with a width of 18 μm (Fig. 5A), a square-shaped track geometry (Fig. 5B), and a circular pattern with a diameter of 40 μm (Fig. 5C). Janus micromotors can reveal a rather complex motion at sophisticated patterns. This behavior is exemplified with motion in a circular ring pattern (channel width: 9 μm; Fig. 5D) and in a square array of circular patterns (diameter of the circle: 5 μm, gap between patterns: 5 μm; Fig. 5E). While the isolated circular patterns are covered with a 5-nm-thick gold layer, the remaining part is a monolayer of APTES. The Janus micromotor can bypass the isolated circular gold patterns by turning the direction all the time. This behavior shows no pinning to the circle's wall previously reported (*18, 19, 21, 34*), and is characterized by multiple reflections of a Janus micromotor from a soft wall.

**Particle transport through the soft wall:** Cell membranes and nanofluidic constrictions enable filtering specific ion species. This effect is not reported at the microscale. We demonstrate the ability to focus and filter Janus particles using flow-focusing constrictions with a gradually changing the ratio of the particle diameter to the constriction width: from 0.14 to 1.1 (Fig. 6, A and B). These geometries successfully focus Janus micromotors towards the narrowest region and guide them through a narrow constriction. To demonstrate the focusing and filtering capabilities, we fabricate flat chemical constrictions with identical dimensions. For this study, we use colloidal Janus particles with different diameters ($d_{JP}$ = 2 μm, 2.39 μm, 3.03 μm, 4 μm). For a constriction of width $W_c$ formed by hard walls, theoretically, almost all Janus particles can pass through the constriction when $d_{JP}/W_c \leq 1$, and the probability becomes 0 for $d_{JP}/W_c > 1$ (Fig. 6C). In the case of charged Janus particles moving in the potential with a soft wall, the Janus particle can pass through the narrow constriction even when $d_{JP}/W_c$ is larger than 1. In particular, Janus particles with the diameter of 4 μm can pass through the narrow focusing part with a width of 3.56 μm (Fig. 6C, $d_{JP}/W_c$ =1.1). Another effect, which is different compared to the case of hard walls, is related to the substantially reduced probability for Janus particles to pass through the constriction for the regime when $d_{JP}/W_c \leq 1$ (Fig. 6C). This is caused by the EOFs generated in the vicinity of the boundaries (Fig. 6D). The focusing geometry with two converging APTES/gold boundaries leads to the slowing down of the Janus particle just before entering the constriction (Fig. 6D). This is due to the fact, that the direction of the EOF on the negatively charged gold surface on both sides is towards the Janus micromotor. Thus, the EOF also plays a significant role in the focusing effect.

Molecular-dynamics simulations (see Methods) of the motion of Janus particles were performed for a double-step repulsive interaction potential consisting of a long-range soft repulsive potential and short-range 'hard-core' potential (Fig. 6E). Employing this type of interaction potential, where the long-range part is equivalent to the exclusion radius (*25, 35*), explains the experimentally observed slowing down of the particle motion when approaching the interface (compare to experimental data, Fig. 4F), the appearance of the potential-energy barrier for a Janus particle when entering a narrow constriction (with a size smaller than the interaction radius) and focusing inside the constriction (compare to experimental data, Fig. 6A).

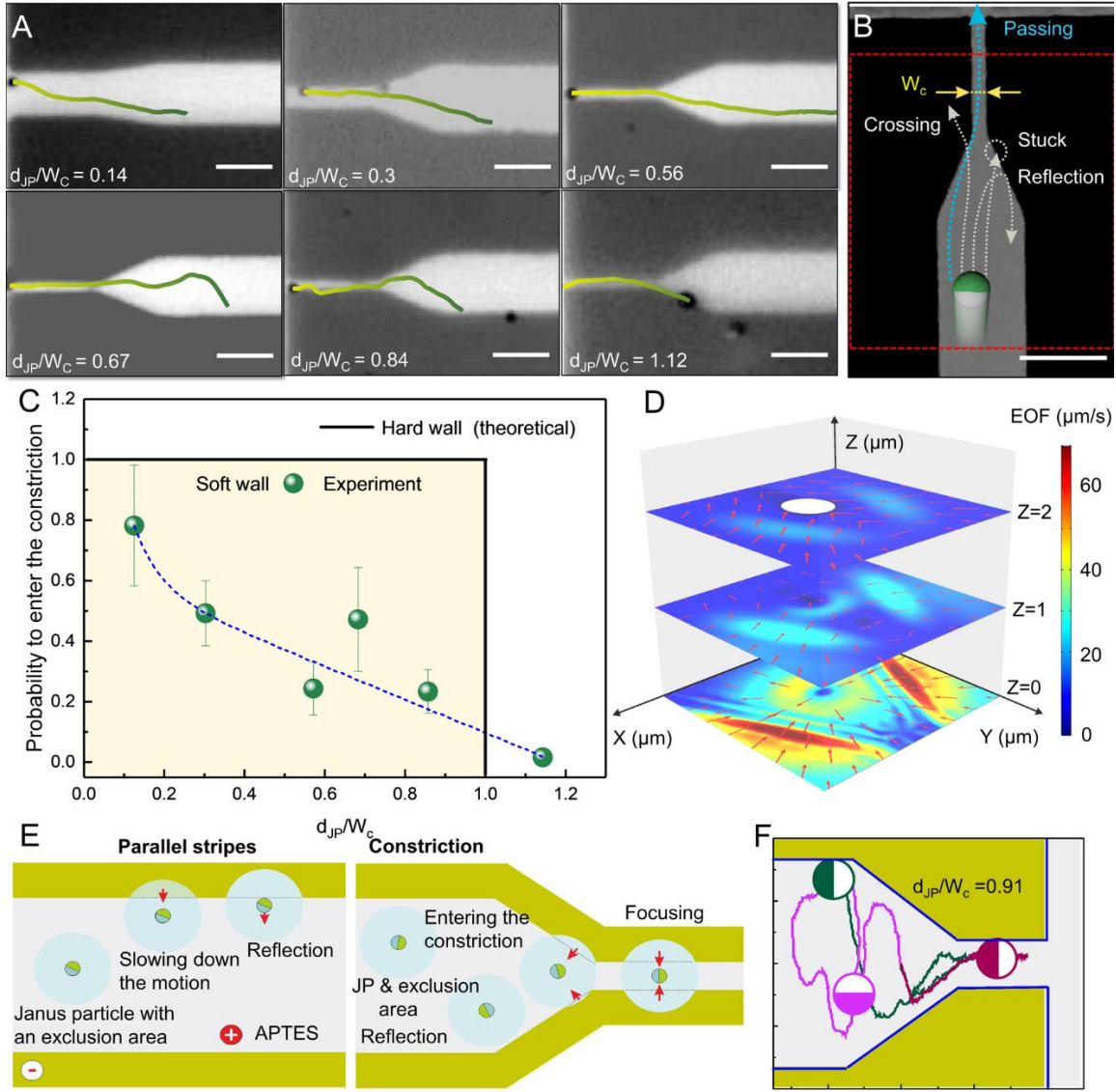

**Figure 6**. (**A**) Experimental trajectories of Janus micromotors moving on the charged pattern with a constriction, for various ratios of $d_{JP}/Wc$. (**B**) A sketch overlaid with a micrograph of the channel with a constriction realized using soft walls with the description of the possible scenario of particle behavior in the constriction: passing (filtering), reflecting, crossing or sticking on the charged interface. (**C**) The probability for a Janus particle to enter a constriction is defined using soft-walls for varying ratio $d_{JP}/Wc$. The theoretical limit for a hard wall channel is indicated with a black frame. The blue dashed line is a guide to the eye to observe the tendency of the probability scaling. (**D**) COMSOL simulations of the electroosmotic flow distribution around a Janus particle on the flow-focusing charged pattern. (**E**) A sketch illustrating an effective repulsive potential created by EOFs around a Janus particle. (**F**) Simulated trajectories of Janus micromotors moving on the charged pattern with a constriction ($d_{JP}/Wc = 0.91$) showing the reflection event and the event when the particle is passing through the constriction. Green trajectory illustrates the scenario when a particle moves back to the wide channel. The pink trajectory shows when a particle moves towards the constriction but is reflected by the step potential barrier. The red trajectory represents when a particle enters the constriction but later turns back.

The resulting trajectories (Fig. 6F) calculated for various ratios $d_{JP}/W_c$ correspond to different scenarios of the particle motion: when a particle moves back to the wide channel, when it moves towards the constriction but is reflected by the step potential barrier (which is specific for the charged pattern system; it is absent in the case of hard-wall constriction), when it enters the constriction but later turns back, and finally when a Janus particle passes all the way through the constriction. Thus, the results of simulations qualitatively reproduce the behavior observed in the experiments including the suppression of the particle velocity near the constriction wall, reflection from the constriction, entering and 'focusing' inside the constriction, and passing through the constriction. The revealed good correspondence between the experimental data and the simulations confirms the relevance of long-range potentials for the description of the behavior of charged Janus particles in channels with soft walls.

**Conclusions**

Inspired by long-range soft potentials for controlling ion-selective transport in nanofluidic systems, we demonstrated the microscale realization of the features typical for the nanoscale ionic transport and validated it for the specific case of electro-kinetic Janus micromotors moving over charged surfaces with a complex spatial charge distribution. We fabricated soft walls represented by an interface between regions of opposite charge, which enable long-range potentials affecting the motion of charged Janus particles. Due to the long-range electrostatic interaction between the charged soft wall and a particle, we observed their unconventional behavior at the boundary and in the constriction: e.g., reflection from the oppositely charged interface, suppressed probability of crossing the topographically flat yet charged interface, flow focusing and filtering. Namely, positively charged self-propelled Janus microspheres exhibited 2D or 3D reflection at the interface formed by the positively/negatively charged regions, or crossing the interface induced by the electro-osmotic flows arising between the Janus micromotors and the charged patterns. By utilizing this reflection behavior, we proposed a novel method to steer Janus motors along complex trajectories by following charged patterns. The revealed phenomena such as particle guidance, flow focusing, and filtering make flat charged patterns promising for future applications in lab-on-chip systems. The demonstrated behavior lays the foundations for developing a new approach to transport species in microfluidics, e.g., performing separation tasks, such as guidance- and size-based sorting. Our results deliver new knowledge on the electro-kinetic transport of biochemical species, as well as on the control of species in fluidic samples in microscale confinement, relevant for the field of ionotronics.

**Methods**

**Materials and Instruments.** The chemicals used, including polyvinylpyrrolidone (PVP) (Mw = 55 000), APTES ((3-Aminopropyl)triethoxysilane), Iron(III) chloride hexahydrate ($FeCl_3 \cdot 6H_2O$), and the polystyrene spheres (nominal diameter: 2 µm) are from Sigma-Aldrich.

**Janus particle fabrication.** Janus particles were fabricated as follows: (*12, 35*) monolayers of polystyrene (PS) spheres with a diameter of 2 µm were prepared by casting a drop of colloidal suspension onto thin glass substrates. Then, a 60-nm-thick Ag layer was thermally evaporated (base pressure $7 \times 10^{-5}$ mbar) onto the PS particles. Afterward, Janus particles were detached from the substrate using an ultrasonication process and resuspended in deionized (DI) water. Particles were further dispersed into a PVP solution (300 mM) for the synthesis of Ag/AgCl layers. The synthesis process of Ag/AgCl was conducted in a dark environment in a solution with an excess concentration of FeCl$_3$ (20 mM) for about 6 h. During this time, β-FeOOH was forming on the surface of the Janus particles by hydrolyzing the FeCl$_3$ solution (*36*). Finally, the colloids with the synthesized Ag/AgCl hemispheres were washed in DI water using a centrifugation process and then dispersed in DI water for further experiments.

**Substrate cleaning.** The glass slides as substrates were pretreated by immersion in a hot H$_2$SO$_4$ : H$_2$O$_2$ (7:3) solution for 30 min, followed by washing with a copious amount of deionized water.

**Charged pattern preparation.** Charged patterns are fabricated with the following process. First, the substrate was functionalized with APTES. The APTES solution was mixed with deionized water and ethanol in the ratio of ethanol : H$_2$O : APTES = 100 : 5 : 2 and left at room temperature for 10 min. Then, the plasma-activated substrate was immersed into the silanization mixture for 1 h with gentle shaking. Next, the substrate was washed three times with pure ethanol and left 15 min at 120°C in an oven. For patterning the substrate, a positive photoresist (S1828 or S1813, Microchemicals, Westborough, USA) was coated at 4000 rpm for 60 s, soft-baked at 115°C for 60 s, and exposed to UV light (Karl Süss MJB4, Garching, Germany). After exposure to UV for 10 s (3.5 s for S1813), the structure is developed for 60 s in AZ 726 MIF (Microresist Technology GmbH, Germany) and rinsed with DI water. Using air plasma for 1 min, silane is removed from the undesired areas. Then the sample is rinsed with DI water. A of 5-nm-thick gold layer is deposited onto the surface via thermal evaporation method (base pressure: 7×10$^{-5}$ mbar). After deposition, the photoresist is removed by sonication in acetone for 10 s followed by a subsequent 30 min washing step in acetone. Lastly, the substrates were washed three times with isopropanol and dried with nitrogen.

**Particle tracking.** TrackMate from the image processing software Fiji (http://fji.sc/) was used to track trajectories of passive beads.

**Simulations.** The dynamics of active Janus particles was simulated by numerically integrating the overdamped Langevin equations (*21, 24, 25, 35, 37, 38*):

$$\dot{x}_i = v_0 \cos\theta_i + \xi_{i0,x}(t) + \sum_{ij}^{N} f_{ij,x},$$
$$\dot{y}_i = v_0 \sin\theta_i + \xi_{i0,y}(t) + \sum_{ij}^{N} f_{ij,y}, \quad (1)$$
$$\dot{\theta}_i = \xi_{i\theta}(t),$$

for *i, j* running from 1 to the total number *N* of particles, $v_0$ is self-velocity of Janus particles which is set to zero for immobilized particles. Here, $\xi_{i0}(t) = (\xi_{i0,x}(t), \xi_{i0,y}(t))$ is a 2D thermal Gaussian noise with correlation functions $\langle \xi_{0,\alpha}(t) \rangle = 0$, $\langle \xi_{0,\alpha}(t)\xi_{0,\beta}(t) \rangle = 2D_T \delta_{\alpha\beta}\delta(t)$, where $\alpha, \beta = x, y$ and $D_T$ is the translational diffusion constant of a passive particle at fixed temperature. $\xi_\theta(t)$ is an independent 1D Gaussian noise with correlation functions $\langle \xi_\theta(t) \rangle = 0$ and $\langle \xi_\theta(t)\xi_\theta(0) \rangle = 2D_R\delta(t)$ that models fluctuations of the propulsion angle $\theta$. The diffusion coefficients $D_T$ and $D_R$ can be directly calculated or extracted from experimentally measured trajectories and MSD, by fitting to theoretical MSD (*25*). Thus, for a particle with diameter of 2 μm diffusing in water at room temperature, $D_T \approx 0.22$ μm$^2$ s$^{-1}$ and $D_R \approx 0.16$ rad$^2$ s$^{-1}$. The term, $\sum_{ij}^{N} f_{ij}$, represents, in a compact form, the sum of all inter-particle interaction forces in the system including (*25, 35*): (i) elastic soft-core repulsive interactions between active particles; here we assume the same short-range interaction between active particles and the interface; and (ii) the effective long-range repulsive interaction between Janus particles and the interface, due to the EOF induced by the Janus particle.

The latter contribution (ii) is modelled by a finite-range field of radial forces, decreasing in amplitude as $1/r$ from the center of a Janus particle: (*25, 35*)

$$f_{ij}^{flow} = \begin{cases} \dfrac{\gamma}{|\vec{r}_i - \vec{r}_j|}, & if \ |\vec{r}_i - \vec{r}_j| > R_i + R_j, \\ 0, & if \ |\vec{r}_i - \vec{r}_j| \gg R_i + R_j, \end{cases} \quad (2)$$

where $\gamma$ is the cumulative "strength of the flow" parameter.

**Acknowledgments**

TH acknowledges the China Scholarship Council (CSC) for financial support. VRM and DM acknowledge support from the Research Foundation-Flanders (FWO-Vl), Grant No. G029322N. AS acknowledges German Research Foundation (DFG), grant SY125/11-1 and SY125/15-1 for financial support. FN is supported in part by: Nippon Telegraph and Telephone Corporation (NTT) Research, the Japan Science and Technology Agency (JST) [via the Quantum Leap Flagship Program (Q-LEAP), the Moonshot R&D Grant Number JPMJMS2061, and the Centers of Research Excellence in Science and Technology (CREST) Grant No. JPMJCR1676], the Japan Society for the Promotion of Science (JSPS) [via the Grants-in-Aid for Scientific Research (KAKENHI) Grant No. JP20H00134], the Army Research Office (ARO) (Grant No. W911NF-18-1-0358), the Asian Office of Aerospace Research and Development (AOARD) (via Grant No. FA2386-20-1-4069), and the Foundational Questions Institute Fund (FQXi) via Grant No. FQXi-IAF19-06. This project is supported in part via the German Research Foundation (DFG) grants BA4986/8−1 and MA5144/14−1.

The authors thank Dr. Markus Löffler (TU Dresden), Ling Ding (Leibniz IFW Dresden) for support with the SEM measurements, Dmitry Belyaev (TU Dresden) and Luis Antonio Panes-Ruiz (TU Dresden) for mask preparation and photolithography, Shadab Anwar (HZDR) for Atomic Force Microscopy measurements. We gratefully acknowledge helpful discussions with Prof. Wei Wang (HIT, Shenzhen) and Dr. Chao Zhou (HIT, Shenzhen).